\documentclass[12pt,twoside,bezier]{article}

\font\eightrm=cmr8
\def\beq{\begin{equation}}
\def\eeq{\end{equation}}
\def\BA{\begin{eqnarray}}
\def\EA{\end{eqnarray}}
\begin{document}


\begingroup
\def\thirteen{\large\bf}
\def\ten{\bf}
\def\eight{\footnotesize}
\baselineskip 15pt
\thispagestyle{empty}

\centerline{\ten Centre de Physique
Th\'eorique\footnote
{\eight Unit\'e Propre de Recherche
7061}, CNRS Luminy, Case 907}

\centerline{\ten F-13288 Marseille -- Cedex 9}

\vskip 2truecm

\centerline{\thirteen ON THE WULFF CONSTRUCTION AS A PROBLEM OF}
\smallskip
\centerline{\thirteen EQUIVALENCE OF STATISTICAL ENSEMBLES}
\bigskip

\centerline{{\bf 
Salvador MIRACLE-SOLE\footnote{\eight Centre de Physique
Th\'eorique, CNRS, Marseille}
{\rm and} Jean RUIZ\footnote{\eight Centre de Physique
Th\'eorique, CNRS, Marseille}}}

\vskip 2truecm

\centerline{\bf Abstract} 

\medskip

In this note, the statistical mechanics of SOS (solid-on-solid)
1-dimen\-sional models under the global constraint 
of having a specified area
between the interface and the horizontal axis, is studied.
We prove the existence of the thermodynamic limits and
the equivalence of the corresponding statistical mechanics.
This gives a simple alternative microscopic proof
of the validity of the Wulff construction for such models,
first established in Ref.\ 5.

\bigskip

Published in {\it On Three Levels:
Micro, Meso and Macroscopic Approaches in Physics,}
M. Fannes and A. Verbeure (Eds.),  
Plenum Press, New York, 1994, pp.\ 295--302.    

\bigskip


 
\bigskip
\bigskip

\noindent Number of figures: 1

\bigskip

\noindent CPT-93/P. 2952

\bigskip

\noindent anonymous ftp : ftp.cpt.univ-mrs.fr

\noindent web : www.cpt.univ-mrs.fr

\endgroup
\newpage


\font\pc=cmcsc10 \rm
\def\relatif{\ \hbox{{\rm Z}\kern-.4em\hbox{\rm Z}}}
\def\reel{I\!\!R}
\def\confh{  \{h\} }
\def\t{\thinspace}
\def\Eq{\eqno}
\def\n{\noindent}

\null
\bigskip
\bigskip

We consider the SOS model defined as follows:
To each site $i$ of the lattice $\relatif$ an integer variable $h_i$
is assigned which represents the height of the interface at this 
site.
The energy 
$ H_N ({\confh}) $
of a configuration 
${\confh} = \{ h_0,h_1,...,h_N \} $,
in the box $0 \le i \le N$, of length $N$,
is equal to the length of the corresponding 
interface
\beq 
H_N ({\confh}) = \sum _{i=1}^N ( 1 + |h_i - h_{i-1}|) 
\label{(1)}    
\eeq
Its weight, at the inverse temperature $\beta $,
is proportional to the Boltzmann factor \  
$ \exp [ - \beta H({\confh})] $.

We introduce the Gibbs ensemble
which consits of all configurations, 
in the box of length $N$,
with specified boundary conditions
$h_0=0$ and $h_N=Y$.
The associated partition function is given by
\beq 
Z_1 (N,Y) = \sum _{\confh} e^{- \beta H({\confh})}
\delta (h_0) \delta (h_N - Y) 
\label{(2)} 
\eeq
where the sum runs over all configurations in the box
and $\delta (t) $ is the dicrete Dirac delta
($\delta (t) = 1 $ if $t=0$ and $\delta (t) = 0 $ 
otherwise).
We define the corresponding free energy per site as
the limit
\beq 
\tau _p (y) = \lim _{N \to \infty } -{ 1 \over {\beta N} }
\ln Z_1 (N,yN) 
\label{(3)} 
\eeq
where $y = - \tan \theta $, the slope of the interface,
is a real number.
This free energy is called the projected surface tension.
The surface tension,
which represents the interfacial free energy per unit length
of the mean interface, 
is
\beq
\tau (\theta ) = \cos \theta \  \tau _p (- \tan \theta ) 
\label{(4)} 
\eeq

We introduce also a second Gibbs ensemble,
which is conjugate to the previous ensemble,
and whose partition function,
in the box of length $N$,
is given by
\beq 
Z_2 (N,x) 
= \sum _{\confh} e^{- \beta H({\confh})}
e^{\beta x h_N} \delta (h_0) 
\label{(5)} 
\eeq
where 
$x \in \reel $
replaces as a thermodynamic parameter the slope $y$
and $h_0=0$. 
We define the associated free energy as
\beq 
\varphi (x) = \lim _{N \to \infty } -{ 1 \over {\beta N} }
\ln Z_2 (N,x) 
\label{(6)} 
\eeq

\medskip

{\bf Theorem 1.}
\it
Limits (3) and (6), which define the above free energies, exist.
The first, $\tau _p$, 
is a positive bounded even convex function of $y$.
The second, $\varphi $,
is a bounded above even concave function of $x$.
Moreover, $\tau _p$ and $ -\varphi $ are conjugate
convex functions, i.e.,
they are related by the Legendre transformations
\BA
- \varphi (x) &=& \sup _y \  [ x y - \tau _p (y) ] \cr
\tau _p (y) &=& \sup _x \  [ x y + \varphi (x) ]  
\label{(7)} 
\EA

\medskip

{\it Proof:}
\rm
The validity of the above statements is well known.
See for instance Refs.\thinspace {2}, {3} 
for a proof of these results
in a more general setting.

\medskip

The convexity of $\tau _p$ is equivalent to the fact that
the surface tension $\tau $ satisfies 
a stability condition called
the triangular inequality 
(see Refs.\ {2}, {3}).
Relations (7) between the free energies express
the thermodynamic equivalence of the two ensembles (2) and (4). 
These relations imply that the curve
$z = \varphi (x)$ gives,
according to the Wulff construction, 
or its modern equivalent the Andreev construction,
the equilibrium shape of the crystal
associated to our system.

The function $\varphi (x)$ defined by (6) is easily computed
by summing a geometrical series. 
One introduces the difference variables
\beq 
n_i = h_{i-1} - h_i 
\label{(8)} 
\eeq
for $i = 1,...,N $, so that
the partition function factorizes
and one obtains
\beq 
\varphi (x) = 1
- \beta ^{-1} \ln \sum _{n\in \relatif } 
e^{ - \beta |n| + \beta x n} 
\label{(9)} 
\eeq
The explicit form of this function is
\beq 
\varphi (x) = 
1
- \beta^{-1} \ln { {\sinh \beta} \over {\cosh \beta - \cosh \beta x} } 
\label{(10)} 
\eeq 
if $-1 < x < 1$ and $ \varphi (x) = - \infty $
otherwise.

\medskip

We next define two new Gibbs ensembles for the system under 
consideration.
In the first of these ensembles we consider the configurations
such that $h_N = 0$, which have  a specified height 
at the origin $h_0 = M$
and which have a specified volume $V$
between the interface and the horizontal axis,
this volume being counted negatively for negative 
heights:
\beq 
V = V({\confh}) 
= \sum _{i=0}^N h_i 
\label{(11)} 
\eeq
The corresponding partition function is
\beq 
Z_3 (N,V,M) = \sum _{\confh} e^{- \beta H({\confh})}
\delta (h_N) \delta (V({\confh}) - V)
\delta (h_0 - M) 
\label{(12)} 
\eeq
The second of these ensembles is the conjugate ensemble of the first
one.
Its partition function is given by
\beq 
Z_4 (N,u,\mu) = \sum _{\confh} e^{- \beta H({\confh})}
e^{\beta u (V({\confh})/N) + \beta \mu h_0 }
\delta (h_N) 
\label{(13)} 
\eeq
where $u \in \reel $ and $\mu \in \reel $
are the conjugate variables.
Our next step will be to prove the existence of the
thermodynamic limit for these ensembles and their
equivalence in this limit.

\medskip

{\bf Theorem 2.} 
\it The following limits exist
\BA
\psi _3 (v,m) 
&=& 
\lim _{N \to \infty } -{ 1 \over {\beta N} }
\  \ln Z_3 (N, v N^2 , mN) 
\label{(14)} 
\\
\psi _4 (u,\mu) &=& \lim _{N \to \infty } -{ 1 \over {\beta N} }
\  \ln Z_4 (N, u,\mu ) 
\label{(15)} 
\EA
They define the free energies per site associated to the considered 
ensembles as, respectively,  
convex and concave functions of the $v$ and $u$ variables.
Moreover, $\psi _3$ and $- \psi _4$ are conjugate convex functions:
\BA
- \psi _4 (u,\mu) 
&=& \sup _{v,m} \  [ u v + \mu m - \psi _3 (v,m) ] 
\cr
 \psi _3 (v, m) 
&=& \sup _{u,\mu} \  [ u v + \mu m + \psi _4 (u,\mu) ] 
\label{(16)} 
\EA
\rm

{\it Proof:}
The crucial observation is the subadditivity property
given in Lemma 1 below.
Then we  addapt
known arguments in the theory of the
thermodynamic limit (see Refs.\ {4}, {5}).
A more detailed discussion is given in the Appendix.

\medskip

{\bf Lemma 1.} 
\it 
The partition function
$Z_3$
satisfy the subadditivity property
\BA 
&& Z_3 ( 2N, 2(V' + V''), M' + M'' ) \cr
&& \ge 
Z_3 ( N, V', M')\  Z_3 ( N, V'', M'') \  
e^{- 2 \beta |M''| / (2N-1) } 
\label{(17)} 
\EA 
\rm

{\it Proof:}
In order to prove this property we
associate a configuration ${\confh}$
of the first system in the box of length $2N$,
to a pair of configurations ${\confh}'$ and ${\confh}''$
of the system in  a box of length $N$, as follows
\BA
h_{2i}
&= h'_{i} + h''_{i} \; , \; i=0,\dots , N
\cr
h_{2i-1} 
& = h'_{i-1 } + h''_{i} \; , \; i=1,\dots , N
\label{(18)} 
\EA
Then 
$h_{2N} = h'_{N} + h''_{N} = 0$,  
$h_{0} = h'_{0} + h''_{0} = M' + M''$ 
and
\begin{eqnarray*}
V({\confh})
&=& 2 \sum_{i=1}^N h'_i 
+ \sum_{i=0}^N h''_i 
+ \sum_{i=1}^N h''_i 
\cr
&=& 2\  [ V({\confh'}) + V({\confh''}) ] - M''. 
\end{eqnarray*}

This shows that the configuration $\confh$ belongs to
$ Z_3 ( 2N, 2(V' + V'') + M'', M' + M'') $.  
Since $ H_{2N}({\confh}) = H_N ({\confh'}) + H_N ({\confh''})  $,
because 
$n_{2i} =n'_i$ 
and 
$n_{2i-1} =n''_i$ 
for $i=1,...,N-1 $, as follows from (18),
we get 
$$
Z_3 ( N, V', M')\  Z_3 ( N, V'', M'') 
\leq 
Z_3 ( 2N, 2(V' + V'') + M'', M' + M'' ).
$$
Then we use the change of variables
$\tilde{h}_i = h_i - [M''/(2N-1)]$
for $i=1,\dots,2N-1$,
$\tilde{h}_0 = h_0$, 
$\tilde{h}_{2N} = h_{2N}=0$ 
which gives
$$
Z_3 ( 2N, V + M'', M )
\leq
e^{ 2 \beta |M''| / (2N-1) }
\ Z_3 ( 2N, V , M )
$$
to conclude the proof.

\medskip

\bf Theorem 3.
\it
The functions $\psi _3$ and $\psi _4$
can be expressed in terms of the functions
$\varphi$ and $\tau_p$
as follows
\BA 
\psi _4 (u,\mu) 
&=&
{1 \over u} \int _0 ^u \varphi (x + \mu) dx 
\label{(19)} 
\\
\psi _3 (v,m) 
&=&
{{1}\over{u_0}}    
\int _{\mu_0} ^{\mu_0 + u_0}  \tau_p (\varphi' (x)) dx 
\label{(20)}
\EA
where $u_0$ and $\mu_0$ satisfy
\label{(21)}\BA 
 {1 \over {u_0^2} } \int _0 ^{u_0} \varphi (x+ \mu_0) dx 
- {1 \over {u_0} } \varphi ( \mu_0 + u_0 ) &=& v 
\\ 
 {1 \over u_0 } [\varphi (\mu_0 ) - \varphi ( \mu_0 + u_0) ]
&=& m
\label{(22)} 
\EA
\rm

{\it Proof:}
We consider again the difference variables
(8) and observe that
$$  
V({\confh}) 
= \sum _{i=0}^N h_i 
= \sum _{i=1}^N \; i n _i 
$$
and therefore
$$ 
Z_4 (N,u,\mu) 
= \prod _{i=1} ^N \Big( \sum _{n_i \in \relatif }
e^{ - \beta |n_i| + \beta ( u / N ) i n_i + \beta \mu n_i} \Big) 
$$
Taking expression (9) into account it follows
$$ 
Z_4 (N,u, \mu ) 
= \exp \Big( - \beta \sum _{i=1} ^N
\varphi \big( { u \over N} i + \mu \big) \Big) 
$$
and
$$ 
\psi _4 (u,\mu)
= 
\lim _{N \to \infty} {1 \over N} 
\sum _{i=1}^N \varphi \big( { u \over N} i + \mu \big) 
=
\lim _{N \to \infty} {1 \over u} 
\sum _{i=1}^N {u \over N} 
\varphi \big( { u \over N} i + \mu \big) 
$$
which implies (19) in the Theorem.

The function $\psi _3$ is determined by the Legendre transform
(16).
The supremum over $u,\mu$ is obtained for the value $ u_0, \mu_0 $
for which the partial derivatives of the right hand side are zero:
$ v + (\partial/ \partial u) \psi _4 ( u_0,\mu_0 ) = 0, \; 
v + (\partial/ \partial \mu) \psi _4 ( u_0,\mu_0 ) = 0  $. 
That is, for $ u_0, \mu_0 $  which satisfy (21).

Then, from (16), (19) and (21), we get
\beq
\psi _3 (v,m) 
= 
2 \psi_4 (u_0, \mu_0)  -
{{1}\over{u_0}}
[(\mu_0 + u_0) \varphi(\mu_0 + u_0)- \mu_0 \varphi(\mu_0)]
\label{(23)} 
\eeq


\setlength{\unitlength}{.7mm}
\begin{center}
\begin{picture}(0,80)

\bezier{200}(-10,50)(0,55)(10,50)
\bezier{200}(10,50)(30,40)(40,20)
\bezier{200}(40,20)(45,10)(50,-10)
\bezier{200}(-10,50)(-30,40)(-40,20)
\bezier{200}(-40,20)(-45,10)(-50,-10)
\put(0,-10){\vector(0,1){70}}
\put(-60,0){\vector(1,0){120}}
\put(10,20){\line(0,1){30}}
\put(10,20){\line(1,0){30}}
\multiput(10,0)(0,2){10}{\line(0,1){1}}
\multiput(40,0)(0,2){10}{\line(0,1){1}}
\put(2,62){$\varphi(x)$}
\put(64,-2){$x$}
\put(8,-6){$\scriptstyle\mu_0$}
\put(34,-6){$\scriptstyle\mu_0+u_0$}
\put(-6,-6){$\scriptstyle{\rm O}$}
\put(4,19){$\scriptstyle{\rm A}$}
\put(42,19){$\scriptstyle{\rm C}$}
\put(12,51){$\scriptstyle{\rm B}$}

\end{picture}
\end{center}

\bigskip\bigskip

\centerline{\eightrm 
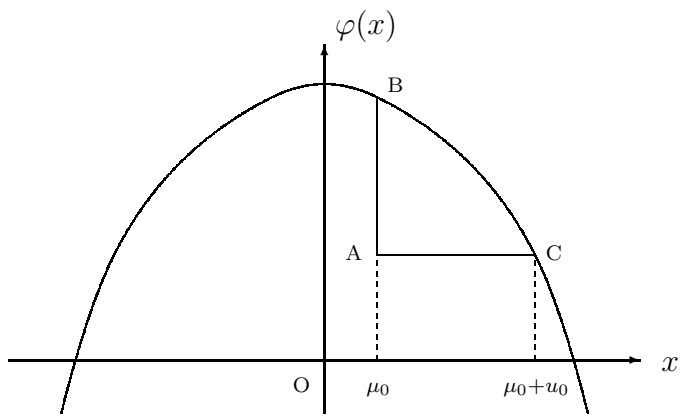
Figure 1. Graphical interpretation of Theorem 3.}

\bigskip\bigskip

\noindent
The right hand side of  (23) represents twice the 
area of the sector $OBC$ in Fig.\thinspace 1 divided by $u_0$.
But, it is a known property 
in the Wulff construction,
that twice this area 
is equal to the integral in (20).
Indeed, by using the relation (7)
in the form
$$ 
\varphi (x) 
= 
x \varphi'(x) + \tau _p (\varphi' (x)) 
$$
in (20) and integrating by parts $x \varphi'(x)$, we get
$$
 2 \psi _4 (u_0,\mu_0) 
=   
{{1}\over{u_0}} 
\int _{\mu_0} ^{\mu_0 + u_0}  \tau_p (\varphi' (x)) dx 
+
{{1}\over{u_0}}
[(\mu_0 + u_0) \varphi(\mu_0 + u_0)- \mu_0 \varphi(\mu_0)]
$$
which together with (23) implies the expression (19) 
in the Theorem.

\medskip

To interpret these relations, let us observe that the right
hand side of (21) represents the area $ABC$, in Fig.\ 1, divided
by $\overline{AC}^2$.
Therefore, the values $u_0$ and $\mu_0$, which solve (21)
and (22), are obtained hen this area is equal to $v$, 
with the condition, coming from (22), that the slope
$\overline{AB}/\overline{AC}$ is equal to $m$.
Then, according to (19), the free energy $\psi_3(v,m)$
is equal to the integral of the surface tension
along the arc $BC$, of the curve $z=\phi(x)$,
divided by the same scaling factor $\overline{AC}=u_0$.

We conclude that, for large $N$, the configurations 
of the SOS model, with a prescribed area $vN^2$,
follow a well defined mean profile,  
the macroscopic profile given by the Wulff construction,
with very small fluctuations.
This follows from the fact that the probability of the 
configurations which deviate macroscopically from the mean 
profile is zero in the thermodynamic limit. 
The free energy associated to the configurations which 
satisfy the above conditions, and moreover, 
are constrained to pass through a given point not 
belonging to the mean profile, 
can be computed with the help of Theorem 3.
The corresponding probabilities decay exponentially 
as $N\to\infty$, as a consequence of the usual
large deviations theory in statistical mechanics
(see Ref.\ {6}).

\bigskip

{\bf Acknowledgements:}
The authors thank Mons University, 
where part of this work was done, 
for warm hospitality and acknowledge the NATO and 
the ERASMUS project for financial support.

\bigskip

\centerline{\bf References}

\begin{enumerate}

\item
J.\t De Coninck, F.\t Dunlop, and V.\t Rivasseau,
{\it 
On the microscopic validity of the Wulff construction
and of the generalied Young equation,}
Commun.\t Math.\t Phys, {\bf 121}, 401--419 (1989)

\item
R.\t L.\t Dobrushin, R.\t Koteck\'y and S.\t B.\t Shlosman,
{\it 
Wulff Construction: A Global Shape from Local Interactions,}
Am.\t Math.\t Soc.\t 
Providence, RI, 
1992.

\item 
A.\t Messager, S.\t Miracle-Sol\'e and J.\t Ruiz, 
{\it
Convexity properties of the surface tension and 
equilibrium crystals,}
J.\t Stat.\t Phys. {\bf 67}, 449--470 (1992)

\item
D.\t Ruelle,
{\it 
Statistical Mechanics, Rigorous Results,}
Benjamin, New York, 1969.

\item
L.\t Galgani, L.\t Manzoni, and A.\t Scotti,
{\it 
Asymptotic equivalence of equilibrium ensembles
of classical statistical mechanics,}
J.\t Math.\t Phys, {\bf 12}, 933--935, (1971)

\item
D.\t Ruelle,
{\it Hazard et Chaos} (chap.\t 19), Odile Jacob, Paris, 1991.
O.\break Lanford,
{Entropy and equilibrium states in classical statistical mechanics},
in:
{\it Statistical Mechanics and Mathematical Problems,}
A.\t Lenard, ed., Springer, Berlin, 1973.

\end{enumerate}

\newpage

\centerline{\bf Appendix}

\medskip

We give here a more detailed discussion of the proof of Theorem 2.
We define:
$$
f_n(v,m)
=
-{ 1 \over {\beta 2^n} }
\ln Z_3 (2^n,  2^{2n} v, 2^n m ).
$$
For $v$ and $m$ of the form $2^{-q}p$, 
the subadditivity property with $N=2^n$, 
$V'=V''=vN^2$ and 
$M'=M''=mN$ 
implies that $f_n$ is
a decreasing sequence: $f_{n+1}(v,m) \leq f_n(v,m)$. Since this sequence is 
bounded from below its limit exits when n tends to infinity. 
Indeed
$$ 
Z_3 (N,V,M) 
\leq 
\sum _{\confh} e^{- \beta H({\confh})}
\delta (h_N)  
$$
The R.H.S. of the above expression is easily computed by inroducing the 
differnce variables (8) and we get that $f_n$ is bounded from below by
$(1/\beta) \ln (1-e^{-\beta})$.
Let us notice that one can obtain a lower bound to $Z_3(N,vN^2,mN)$ 
by restricting
the summation over the configuration such that 
$(N-1)h_i = vN^2-(1/2)mN$ for $i=1,\dots,N-1$.
This gives that $f_n$ is bounded from above by
$|v-m| + |v|$.

The existence 
of the limit for general $N$ follows from standard argument in the theory 
of the thermodynamic limit (cf.\thinspace {4}) and we have:
$$
\psi_3(v,m) =
\inf _N \left[ -{ 1 \over {\beta N} } \ln Z_3 (N, v N^2,mN )\right]
\Eq{(A.1)}
$$

To prove that $\psi_3$ is convex, we notice that the 
subadditivity inequality (17) with 
$N=2^n$, $V'=v_1 N^2$,  $V''=v_2 N^2$ and $M'=M''=mN^2$ gives:
$$\psi_3 \Big({1\over 2} (v_1 +v_2,m)\Big)\leq 
{1\over 2} \psi_3(v_1,m)
+ {1\over 2} \psi_3(v_2,m)  
$$
which applied iteratively implies:
$$
\psi_3 \Big(\alpha v_1 + (1- \alpha) v_2,m \Big)\leq 
\alpha  \psi_3(v_1,m)
+ (1 -\alpha) \psi_3(v_2,m)
$$
for $\alpha$ of the form $2^{-q}p$ and $0\leq \alpha \leq 1$.
For such $\alpha$ we obtain analogously
$$
\psi_3 \Big( v ,  \alpha m_1 + (1- \alpha) m_2 \Big)\leq 
\alpha  \psi_3(v,m_1)
+ (1 -\alpha) \psi_3(v,m_2)
$$
by applying the subadditivity inequality (17) with
$N=2^n$, $M'=m_1 N^2$,  $M''=m_2 N^2$ and $V'=V''=vN^2$

Since $\psi _3$ is bounded, it follows that for all $m$, 
$\psi _3$ can be extended to a convex Lipshitz continuous
function of the real variable $v$.

To prove the  the existence of the  limit (15)
and relations (16),
we introduce 
$$
Z_4^+(N,u,\mu)
=\sup_{V,M \in \relatif } 
\left[ e^{\beta u  (V/N) + \beta \mu M } 
\  Z_3(N,V,M)
 \right]  
$$
and proceed, as in the Appendix of Ref.\thinspace {5}, 
to study the thermodynamic limit for this quantity.
Let
$$
\psi^{\ast}_4 (u,\mu) = \sup_{v,m} \  [uv+\mu m -\psi_3 (v,m)] .
$$	
According to (A.1), we 
have 
$$
e^{\beta u (V/N) + \beta \mu M} \  Z_3(N,V,M)
\leq  
e^{\beta N [uv+\mu m - \psi_3(v,m)]}
$$
for all 
$ V \in \relatif$,
so that
$$
Z_4^+(N,u,\mu) \;
\leq \; 
e^{\beta N \psi_4^{\ast}(u)}.
a\Eq{(A.2)}
$$
On the other hand for any 
$\delta>0$ 
and sufficiently large $N$ one can find,
$V = v N^2$ and $M=mN$ 
such that 
\begin{eqnarray*}
e^{\beta u (V/N)+\beta \mu M} \;  Z_3(N,V,M)
&=& e^{\beta N [uv+\mu m - \psi_3(v,m)]}
\; e^{\beta N \psi_3(v,m)}
\; Z_3(N,V,M)
\cr 
&\geq& 
e^{\beta N [\psi_4^{\ast}(u,\mu)- \delta]}
\end{eqnarray*}
and 
therefore
$$
Z_4^+(N,u,\mu) \;
\geq \;
e^{\beta N [\psi_4^{\ast}(u,\mu)- \delta]} . 
\Eq{(A.3)}
$$
Inequalities 
(A.2) and (A.3) imply that 
$$
\lim_{N \rightarrow \infty}\; 
-{1\over \beta N} \; \ln Z_4^+(N,u,\mu) \;
=\; - \psi_4^{\ast}(u,\mu)\; . 
\Eq{(A.4)}
$$
We shall now prove that the thermodynamic limit (15) exits 
and gives the same 
quantity  following the argument of Theorem 2
in Ref.\thinspace {5}. 
First, we notice that 
$$ 
Z_4(N,u,\mu)
=\sum_{V,M \in \relatif }  
e^{\beta u  (V/N) + \beta \mu M } \  Z_3(N,V,M)     
$$
which implies 
$$
Z_4^+(N,u,\mu)
\leq 
Z_4(N,u,\mu)
\Eq{(A.5)} 
$$
Moreover, the inequality
$$
Z_3(N,V,M)  
\leq 
e^{-\beta {\bar u} (V/N)-\beta {\bar {\mu}} M} 
\; Z_4^+(N,{\bar u},{\bar {\mu}}) 
$$
used with ${\bar u}=u'$ and ${\bar u}=u''$,
${\bar {\mu}}=\mu'$ and ${\bar {\mu }}=\mu''$ 
gives for any $u''<u<u'$ and any $\mu''<\mu < \mu'$ :
$$
\begin{array}{rcl}
Z_4(N,u,\mu) 
&\leq& 
Z^{\ast}
\sum_{V\geq 0}  e^{\beta (u-u')(V/N)}
\sum_{M\geq 0} e^{\beta (\mu-\mu')M}
\cr
&+&
Z^{\ast}
\sum_{V\leq 0}  e^{\beta (u''-u)(V/N)}
\sum_{M\geq 0} e^{\beta (\mu-\mu')M}
\cr
&+&
Z^{\ast}
\sum_{V\geq 0}  e^{\beta (u-u')(V/N)}
\sum_{M\leq 0} e^{\beta (\mu''-\mu)M}
\cr
&+&
Z^{\ast}
\sum_{V\leq 0}  e^{\beta (u''-u)(V/N)}
\sum_{M\leq 0} e^{\beta (\mu''-\mu)M}
\cr
&\leq&
{{  4 Z^{\ast}  }\over{  (1-e^{-\beta (\Delta u/N)})}  
(1-e^{-\beta (\Delta \mu)}) }
\end{array}
\Eq{(A.6)}
$$     
where 
$$
\begin{array}{rcl}
Z^{\ast}
&=&\sup 
\big[
 Z_4^+(N,u',\mu'), 
Z_4^+(N,u',\mu'') , 
Z_4^+(N,u'',\mu') , 
Z_4^+(N,u'',\mu'')  
\big]
\cr  
0< \Delta u &=& \min [u'-u,u-u'']
\cr
0< \Delta u &=& \min [\mu'-\mu,\mu-\mu'']
\cr
\end{array}
$$
By referring to (A.4), and to the continuity of $\psi^{\ast}_4$, the  
inequalities (A.5) and (A.6) imply (15) and (16).

\end{document}